\documentclass[twocolumn,showpacs,preprintnumbers,amsmath,amssymb]{revtex4-1}  
\usepackage{graphicx}
\usepackage{dcolumn}
\usepackage{bm}

\begin{document}

\title{JLab Measurement of the $^4$He Charge Form Factor
at Large Momentum Transfers}

\author
{
{A.~Camsonne}$^{22}$,
{A.T.~Katramatou}$^{11}$,
{M.~Olson}$^{20}$,
{N.~Sparveris}$^{11,21}$,
{A.~Acha}$^{4}$,
{K.~Allada}$^{12}$,
{B.D.~Anderson}$^{11}$,
{J.~Arrington}$^{1}$,
{A.~Baldwin}$^{11}$,
{J.-P.~Chen}$^{22}$,
{S.~Choi}$^{18}$,
{E.~Chudakov}$^{22}$,
{E.~Cisbani}$^{8,10}$,
{B.~Craver}$^{23}$,
{P.~Decowski}$^{19}$,
{C.~Dutta}$^{12}$,
{E.~Folts}$^{22}$,
{S.~Frullani}$^{8,10}$,
{F.~Garibaldi}$^{8,10}$,
{R.~Gilman}$^{16,22}$,
{J.~Gomez}$^{22}$,
{B.~Hahn}$^{24}$,
{J.-O.~Hansen}$^{22}$,
{D.~Higinbotham}$^{22}$,
{T.~Holmstrom}$^{13}$,
{J.~Huang}$^{14}$,
{M.~Iodice}$^{9}$,
{X.~Jiang}$^{16}$,
{A.~Kelleher}$^{24}$,
{E.~Khrosinkova}$^{11}$,
{A.~Kievsky}$^{7}$,
{E.~Kuchina}$^{16}$,
{G.~Kumbartzki}$^{16}$,
{B.~Lee}$^{18}$,
{J.J.~LeRose}$^{22}$,
{R.A.~Lindgren}$^{23}$,
{G.~Lott}$^{22}$,
{H.~Lu}$^{17}$,
{L.E.~Marcucci}$^{7,15}$,
{D.J.~Margaziotis}$^{2}$,
{P.~Markowitz}$^{4}$,
{S.~Marrone}$^{6}$,
{D.~Meekins}$^{22}$,
{Z.-E.~Meziani}$^{21}$,
{R.~Michaels}$^{22}$,
{B.~Moffit}$^{14}$,
{B.~Norum}$^{23}$,
{G.G.~Petratos}$^{11}$,
{A.~Puckett}$^{14}$,
{X.~Qian}$^{3}$,
{O.~Rondon}$^{23}$,
{A.~Saha}$^{22}$,
{B.~Sawatzky}$^{21}$,
{J.~Segal}$^{22}$,
{M.~Shabestari}$^{23}$,
{A.~Shahinyan}$^{25}$,
{P.~Solvignon}$^{1}$,
{R.R.~Subedi}$^{23}$,
{R.~Suleiman}$^{22}$,
{V.~Sulkosky}$^{22}$,
{G.M.~Urciuoli}$^{8}$,
{M.~Viviani}$^{7}$,
{Y.~Wang}$^{5}$,
{B.B.~Wojtsekhowski}$^{22}$,
{X.~Yan}$^{18}$,
{H.~Yao}$^{21}$,
{W.-M.~Zhang}$^{11}$,
{X.~Zheng}$^{23}$,
{L.~Zhu}$^{5}$
}


\affiliation{$^{1}$Argonne National Laboratory, Argonne, IL 60439, USA}
\affiliation{$^{2}$California State University, Los Angeles, CA 90032, USA }
\affiliation{$^{3}$Duke University (TUNL), Durham, NC 27708, USA}
\affiliation{$^{4}$Florida International University, Miami, FL 33199, USA}
\affiliation{$^{5}$University of Illinois at Urbana Champagne, Urbana, IL 61801, USA}
\affiliation{$^{6}$Istituto Nazionale di Fisica Nucleare, Sezione di Bari and University of Bari, 70126 Bari, Italy}
\affiliation{$^{7}$Istituto Nazionale di Fisica Nucleare, Sezione di Pisa, 56127 Pisa, Italy}
\affiliation{$^{8}$Istituto Nazionale di Fisica Nucleare, Sezione di Roma, 00185 Rome, Italy}
\affiliation{$^{9}$Istituto Nazionale di Fisica Nucleare, Sezione di Roma Tre, 00146 Rome, Italy}
\affiliation{$^{10}$Istituto Superiore di Sanit\`{a}, 00161 Rome, Italy}
\affiliation{$^{11}$Kent State University, Kent OH 44242, USA}
\affiliation{$^{12}$University of Kentucky, Lexington, KY 40506, USA}
\affiliation{$^{13}$Longwood University, Farmville, VA 23909, USA}
\affiliation{$^{14}$Massachusetts Institute of Technology, Cambridge, MA 02139, USA}
\affiliation{$^{15}$University of Pisa, 56127 Pisa, Italy}
\affiliation{$^{16}$Rutgers, The State University of New Jersey, Piscataway, NJ 08855, USA}
\affiliation{$^{17}$University of Science and Technology of China, Hefei, Anhui, 230026 P.R. China}
\affiliation{$^{18}$Seoul National University, Seoul 151-747, Korea}
\affiliation{$^{19}$Smith College, Northampton, MA 01063, USA}
\affiliation{$^{20}$St.~Norbert College, De Pere WI 54115, USA}
\affiliation{$^{21}$Temple University, Philadelphia, PA 19122, USA}
\affiliation{$^{22}$Thomas Jefferson National Accelerator Facility, Newport News, VA 23606, USA}
\affiliation{$^{23}$University of Virginia, Charlottesville, VA 22901, USA}
\affiliation{$^{24}$College of William and Mary, Williamsburg, VA 23185, USA}
\affiliation{$^{25}$Yerevan Physics Institute, Yerevan 375036, Armenia}


\date{September 10, 2013; The Jefferson Lab Hall A Collaboration}

\begin{abstract}

The charge form factor of $^4$He has been extracted in the range
29 fm$^{-2}$ $\le Q^2 \le 77$ fm$^{-2}$ from elastic electron scattering,
detecting $^4$He nuclei and electrons in coincidence with the High Resolution
Spectrometers of the Hall A Facility of Jefferson  Lab.  The results are in
qualitative agreement with realistic meson-nucleon theoretical calculations.
The data have uncovered a second diffraction minimum, which was predicted in the $Q^2$
range of this experiment, and rule out conclusively long-standing predictions of
dimensional scaling of high-energy amplitudes using quark counting.

\vspace* {-.1in}

\end{abstract}

\pacs{25.30.Bf, 13.40.Gp, 27.10.+h, 24.85.+p}

\maketitle

The electromagnetic (EM) form factors of the helium isotopes
are, along with the deuteron and tritium form factors,
the ``observables of choice''~\cite{ma98} for testing the nucleon-meson
standard model of the nuclear  interaction and the associated EM current
operator~\cite{ca98}.  They provide fundamental
information on the internal structure and dynamics of the light nuclei
as they are, in a simple picture, convolutions of the ground
state wave function
with the EM form factors of the constituent nucleons.
The theoretical calculations for these few-body observables are very
sensitive to the model used for the nuclear EM current operator, especially
its meson-exchange-current (MEC) contributions.  Relativistic corrections
and possible admixtures of multi-quark states in the nuclear wave
function might also be relevant~\cite{ca98}.  Additionally, at large momentum
transfers, these EM form factors
may offer a unique opportunity to uncover a possible
transition in the description of elastic electron scattering on
few-body nuclear systems, from meson-nucleon to quark-gluon degrees of
freedom, as predicted by quark-dimensional scaling~\cite{br76,ca97}.

Experimentally, the few-body form factors are determined
from elastic electron-nucleus scattering studies using high intensity
beams, high density targets and large solid angle magnetic spectrometers.
There have been extensive experimental investigations
of the few-body form factors over the past 50 years at almost every electron
accelerator laboratory~\cite{si01,gi02}, complemented by equally extensive theoretical
calculations and predictions~\cite{ca98,gi02}.

This work focuses on a measurement of the $^4$He charge form factor, $F_c$, at
large momentum transfers, at Jefferson Lab (JLab).
The cross section for elastic scattering of a relativistic electron
from the spin zero $^4$He  nucleus is given, in the one-photon (between
electron and nucleus) exchange approximation, by the formula~\cite{dr64}:
\begin{equation}
{ {d\sigma} \over {d\Omega} } (E,\theta)  =
{ {(Z\alpha)^2 E^\prime \cos^2 \left( {\theta \over 2} \right)}
\over {4 E^3  \sin^4 \left( {\theta \over 2}\right)} }
 {F^2_c(Q^2) },
\label{roseform}
\end{equation}
where $\alpha$ is the fine-structure constant, $Z$ is the nuclear charge,
$E$ and $E'$ are the incident and scattered electron energies,
$\theta$ is the electron scattering angle, and $Q^2 = 4 E E' \sin^2 (\theta/2)$
is the four-momentum transfer squared.

The few-body EM form factors have been theoretically
investigated by several groups, using different techniques to solve
for the nuclear ground states, and a variety of models for the nuclear
EM current. The most recent calculation of $^3$H and $^3$He EM
form factors is that of Refs.~\cite{ma98,ma05}.  It uses the pair-correlated
hyperspherical harmonics (HH) method~\cite{ki08} to obtain the few-body
nuclear wave functions and goes beyond the impulse approximation (IA),
where the electron interacts with one of the nucleon constituents,
by including MEC, whose main contributions are constructed to satisfy
the current conservation relation with the given Hamiltonian~\cite{ma05}.
Part of the present work is the extension of the above method to the $^4$He
charge form factor (see Figs. 1 and 2) by using the (uncorrelated)
HH expansion to solve for the $^4$He wave function from the
Argonne AV18~\cite{wi95} nucleon-nucleon ($NN$) and
Urbana IX~\cite{pu95} three-nucleon ($3N$) interactions,
and including MEC contributions arising from $\pi$-, $\rho$-
and $\omega$-meson exchanges, as well as the $\rho\pi\gamma$ and
$\omega\pi\gamma$ charge transition couplings. For more details,
the reader is referred to Ref.~\cite{ki08} for the HH method, and
Refs.~\cite{ma98,ma05} for the nuclear EM current model. The present
experimental and theoretical results are compared to (see below) (i) the Monte Carlo
calculations of Refs.~\cite{sc90,wi91}, where the
variational Monte Carlo (VMC) and the
Green's function Monte Carlo (GFMC) methods were used to solve for the
$^4$He wave function, and (ii) the long-standing prediction of the
dimensional-scaling quark model (DSQM) approach of Ref.~\cite{br76}.

In fact, at large momentum transfers, elastic scattering from few-body
nuclear systems like $^4$He may be partly due to, or even dominated
by, contributions from electron interaction with the nucleons'
constituent quarks.  Several groups have developed purely
phenomenological ``hybrid quark-hadron'' models that include
multi-quark states for overlapping nucleons in the nuclear
wave function, which augment the IA calculated form factors~\cite{na82}.
On the other hand, the DSQM approach incorporates
the nucleon's quark-gluon substructure in elastic electron scattering
from few-body systems by applying
dimensional scaling of high energy amplitudes using
quark counting. This leads to the prediction for the $^4$He case
that $\sqrt{F_c(Q^2)} \sim (Q^2)^{1-3A}$, where $A$~=~4, and to
the dominance of the constituent-interchange force between quarks
of different nucleons to share $Q/A$ (see Ref. \cite{br76}).

The experiment (E04-018) used the Continuous Electron Beam (100$\%$ duty factor)
Accelerator and Hall A Facilities of JLab.
Electrons scattered from a high density cryogenic $^4$He target were
detected in the Left High Resolution Spectrometer (e-HRS).  To
suppress backgrounds and unambiguously separate elastic from inelastic
processes, recoil helium nuclei were detected in the Right HRS (h-HRS)
in coincidence with the scattered electrons.

The energy of the incident beam ranged between
2.09 and 4.13 GeV.  The beam current was measured using two resonant
cavity current monitors upstream of the target.  It ranged, on average for different
kinematical settings, between 38 and 82$\mu$A.
The two cavities were calibrated against a
parametric current transformer monitor (Unser monitor).
To reduce beam-induced
target density changes and to avoid possible destruction of the target cell,
the beam was rastered on the target in both horizontal and vertical
directions at high frequency, resulting in an effective beam spot size of
$2 \times 2$ mm$^2$.

The target system contained gaseous $^4$He and liquid hydrogen
cells of length $T$=20~cm.  The $^4$He gas was pressurized to 13.7-14.2~atm at
a temperature of 7.14-8.68~K, resulting in a density of 0.102-0.127 g/cm$^3$.
Two Al foils separated by 20~cm
were used to measure any possible contribution to the cross section from the
Al end-caps of the target cells.  This system provided, at the maximum beam
current of 110~$\mu$A, a record high luminosity of
$2.7 \times 10^{38}$ cm$^{-2}$s$^{-1}$, for the $^4$He target.


Scattered electrons were detected in the e-HRS using two planes of
scintillators to form an ``electron'' trigger, a pair of drift
chambers for electron track reconstruction, and a gas threshold
\v{C}erenkov counter and a lead-glass calorimeter for electron
identification.  Recoil nuclei were detected in the h-HRS
using two planes of scintillators to form a ``recoil'' trigger
and a pair of drift chambers for recoil track reconstruction.
The event trigger consisted of a coincidence between the two HRS triggers.
Details on the Hall A Facility and all associated beam, target and spectrometer
apparatuses used are given in Ref. \cite{nimA}.

\begin{figure} [t]
\includegraphics[width=68mm, angle=90]{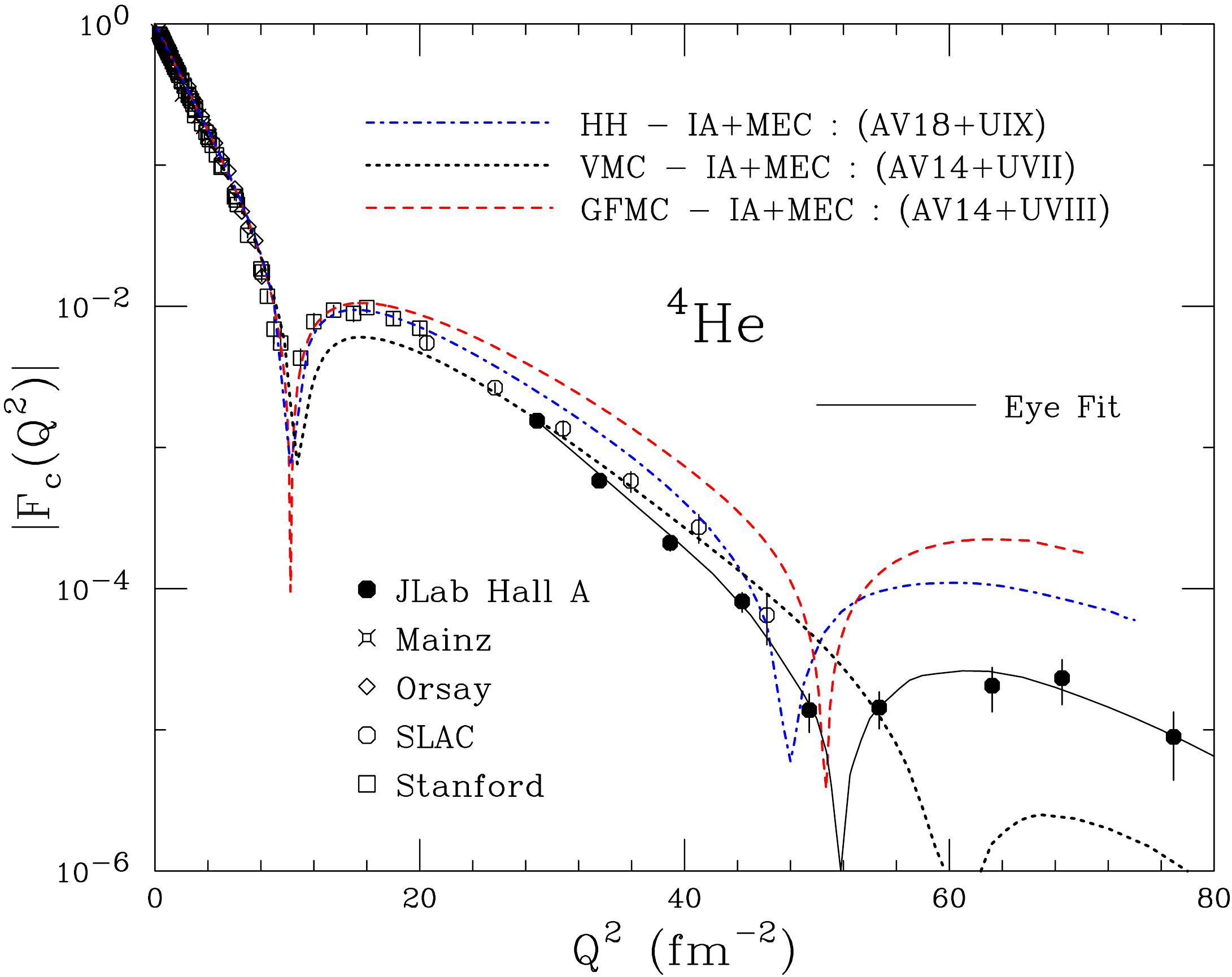}
\caption{\label{fig:fciamec} $^4$He charge form factor data
from this experiment are compared with the present HH theoretical
IA+MEC calculation
using the AV18+Urbana IX Hamiltonian model. Also shown are
previous Stanford, Orsay, Mainz and SLAC data, and older
VMC and GFMC theoretical calculations (see text).  The solid line has been
drawn to just guide the eye.}
\vspace* {-.2in}
\end{figure}


Particles in the e-HRS were identified as electrons on the basis of
a minimal pulse height in the \v{C}erenkov counter
(``\v{C}erenkov cut'')
and the energy deposited in the calorimeter, consistent with the
momentum as determined from the drift chamber track using the spectrometer's
optical properties
(``calorimeter cut'').
Particles in the h-HRS were identified as $^4$He on the basis of their
energy deposition (pulse height) in the first scintillator hodoscope
(``helium cut'').
Electron-$^4$He~($e$-$^4He$) coincidence events
were identified using the relative time-of-flight (TOF) between
the electron and recoil triggers after imposing the above three cuts.
To check the overall normalization, elastic
electron-proton ($e$-$p$) scattering was measured at several kinematics
with solid angle Jacobians similar to those for $e$-$^4He$ elastic scattering.
The $e$-$p$ measured cross section values were found to be in
excellent agreement (to within $\pm2.0\%$) with values
calculated using a proton form factor fit~\cite{jona} based on all
existing $e$-$p$ elastic cross section measurements.

The elastic $e$-$^4He$ cross section values were calculated using the formula:
\begin{equation}
\frac {d\sigma} {d\Omega} (E,\theta)=\frac {N_{er}C_{cor}}
{N_b N_t (\Delta\Omega)_{MC} F(Q^2,T)},
\end{equation}
where $N_{er}$ is the number of electron-recoil $^4$He elastic events,
$N_b$ is the number of incident beam electrons, $N_t$ is the
number of target nuclei/cm$^2$, $(\Delta\Omega)_{MC}$ is the
effective coincidence solid angle (which includes most radiative effects)
from a Monte Carlo simulation,
$F$ is the portion of the radiative corrections that depends only on $Q^2$
and $T$ (1.10 on average)~\cite{moca},
and $C_{cor}=C_{det}C_{cdt}C_{rni}C_{den}$.  Here, $C_{det}$ is the
correction for the inefficiency of the \v{C}erenkov counter
and the calorimeter (1.01$\%$) (the scintillator counter hodoscopes and
the drift chamber sets were found to be essentially 100$\%$ efficient),
$C_{cdt}$ is the computer dead-time correction (between 1.05 and 1.17),
$C_{rni}$ is a correction for losses of recoil nuclei due to
nuclear interactions in the target cell and vacuum windows [1.10(1.03)
at the lowest(highest) $Q^2$], and $C_{den}$ is a correction to
the target density due to beam heating effects
(ranging between 1.03 at 38$\mu$A and 1.06 at 82$\mu$A).
There were no contributions to the elastic $e$-$^4He$ cross section from events
originating in the target cell end-caps, as determined from runs with the
empty replica target.  The $e$-$p$ elastic cross section values were
determined similarly.

\begin{figure}[b]
\includegraphics[width=68mm, angle=90]{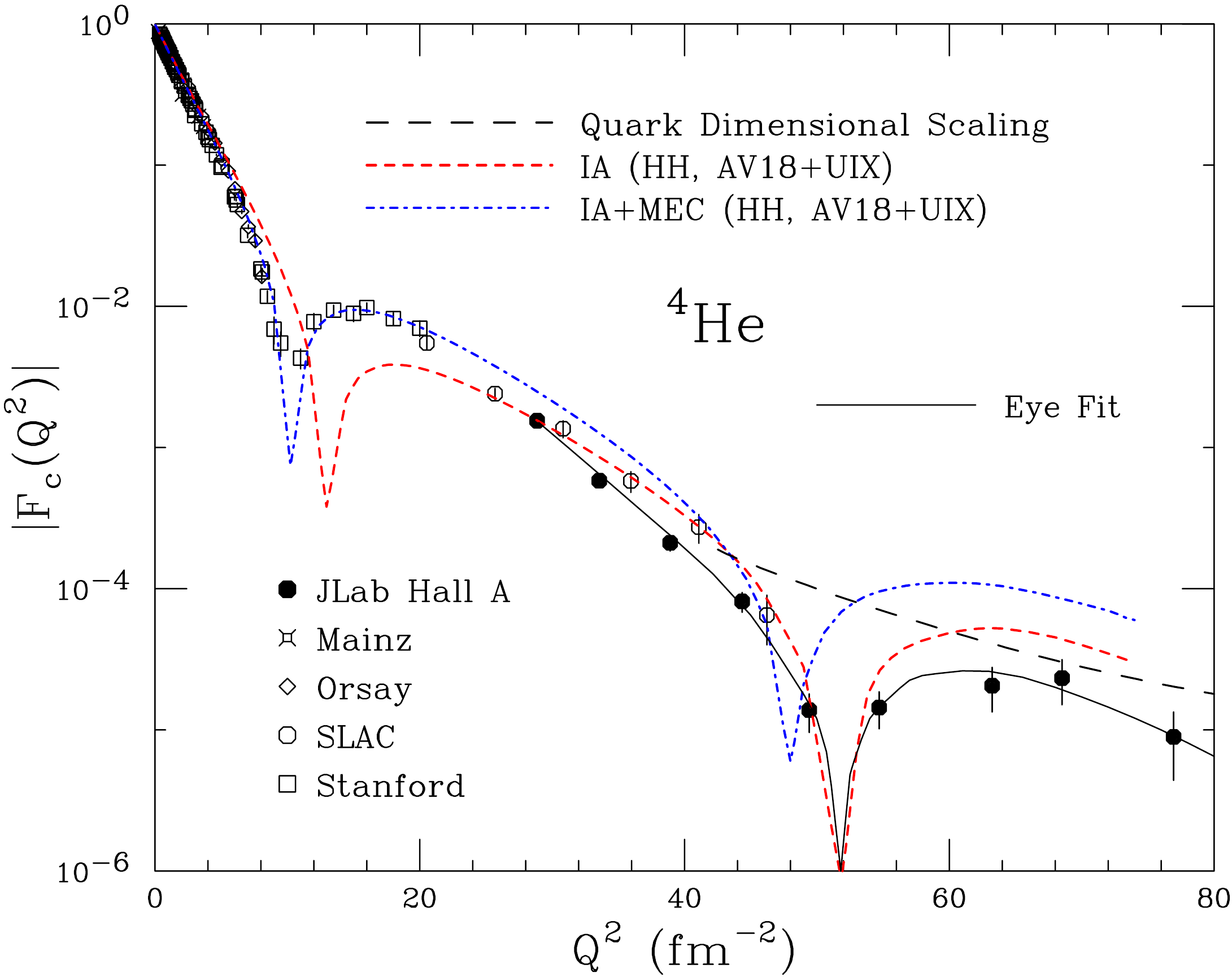}
\caption{\label{fig:fcia}
$^4$He charge form factor data from this experiment
are compared to both IA and IA+MEC HH present calculations,
which use the AV18+Urbana IX Hamiltonian model, and
to the DSQM prediction (see text).
Also shown are
the previous Stanford, Orsay, Mainz and SLAC data.}
\vspace* {-.3in}
\end{figure}

The effective coincidence solid angle was evaluated with a
Monte Carlo computer code that simulated elastic
electron-nucleus scattering under identical conditions as our
measurements.  The code tracked scattered electrons and
recoil nuclei from the target to the detectors through the
two HRS systems using optical models based on magnetic field
measurements and precision position surveys of their elements.
The effects from ionization energy losses
and multiple scattering in the target and vacuum windows were
taken into account for both electrons and recoil nuclei.
Bremsstrahlung radiation losses
for both incident and scattered electrons in the target and
vacuum windows, as well as internal radiative effects, were also
taken into account.  Details on this simulation method can
be found in Ref. \cite{moca}.
Monte Carlo simulated spectra of scattered electrons and recoil
nuclei were found to be in very good agreement with experimentally
measured spectra.


\begin{table}
\begin{tabular}{ccccc}
\hline
$Q^2$     & $E$ & $\theta$ & $d\sigma/d\Omega$ & $|F_c|$ \rule[0mm]{0mm}{3.5mm} \\
fm$^{-2}$ & GeV & deg.  & cm$^2$/sr  &   \\
\hline
28.87&2.091&30.52& $(2.04\pm0.18)\times10^{-36}$& $(1.55\pm0.07)\times10^{-3}$ \rule[0mm]{0mm}{3.5mm} \\
33.56&2.091&33.20& $(1.99\pm0.22)\times10^{-37}$& $(5.77\pm0.32)\times10^{-4}$  \\
38.92&2.091&36.11& $(1.69\pm0.42)\times10^{-39}$& $(2.01\pm0.23)\times10^{-4}$  \\
44.36&4.048&19.25& $(9.51\pm2.76)\times10^{-39}$& $(8.01\pm0.12)\times10^{-5}$  \\
49.43&4.048&20.40& $(2.14\pm1.01)\times10^{-40}$& $(1.36\pm0.32)\times10^{-5}$  \\
54.71&4.048&21.56& $(1.87\pm0.88)\times10^{-40}$& $(1.42\pm0.33)\times10^{-5}$  \\
63.23&4.127&22.86& $(2.84\pm1.91)\times10^{-40}$& $(2.02\pm0.68)\times10^{-5}$  \\
68.51&4.127&23.90& $(2.97\pm1.99)\times10^{-40}$& $(2.26\pm0.76)\times10^{-5}$  \\
76.95&4.127&25.50& $(3.31\pm3.38)\times10^{-41}$& $(8.67\pm4.43)\times10^{-6}$  \\
\hline
\end{tabular}
\caption{Kinematics, elastic $e$-$^4He$ cross section and $^4$He charge
form factor results from this experiment, and total errors (statistical
and systematic added in quadrature).}
\label{data}
\vspace* {-.1in}
\end{table}

The extracted $^4$He charge form factor (absolute) values are listed in
Table~\ref{data},
and shown in Fig.~\ref{fig:fciamec} along with previous Stanford~\cite{hepl},
Orsay~\cite{re65}, SLAC~\cite{ar78} and Mainz~\cite{ot85} data.
The error bars in Fig.~\ref{fig:fciamec} represent statistical and systematic
uncertainties added in quadrature.
The solid curve in Fig.~\ref{fig:fciamec} labeled as ``eye fit"is a line
drawn just to guide the eye.  The new data in the figure suggest
the existence of
a second diffraction minimum for the $^4$He form factor at
$Q^2$~=~(51.7 $\pm$ 0.2)~fm$^{-2}$.
The existence of the minimum is confirmed by
the momentum distribution of the observed $e$-$^4He$ elastic events for
the two $Q^2$ points about the minimum, 50 and 55 fm$^{-2}$: for the
former~(latter) point, the distribution is indicative of a fast
falling~(rising) form factor with $Q^2$.
It is also evident from Fig. 1 that the new JLab data are
in significant disagreement with the existing SLAC data.


The data in Fig.~\ref{fig:fciamec} are compared to the HH variational
calculation performed using the AV18 $NN$ and Urbana IX $3N$ interactions.
Also shown are the VMC results of Ref.~\cite{sc90}, obtained with the
older Argonne AV14 $NN$ and Urbana VII $3N$ interaction, and
the GFMC results of Ref.~\cite{wi91}, obtained with the AV14 and
Urbana VIII $3N$ force model.  It can be seen that all three calculations,
which include MEC contributions, are in qualitative agreement
with the new JLab data and do predict, though at different locations,
a second diffraction minimum
for $Q^2 > 40$ fm$^{-2}$.
The present HH calculation for the $^4$He $F_c$ is in a qualitatively
better agreement with the data when compared with the older
Monte Carlo studies of Refs.~\cite{sc90,wi91}. To better investigate
this aspect, we show in Fig.~\ref{fig:fcia} the experimental data
along with the HH results, with and without (IA only) inclusion of MEC.
Of note is that the lower $Q^2$ data are in good agreement with the
calculation that includes MEC, while the higher $Q^2$ data are in better
agreement with the IA only calculation.  This observation may be indicative
of a possible diminishing role of MEC with increasing $Q^2$ required to bring
the theory into better agreement with the data.  The inadequacy of the
above theoretical approach to describe well the entire $Q^2$ range
 of the $^4$He $F_c$ may
also indicate the need for a truly covariant relativistic framework, which
has been successful in describing all deuteron form factor
data~\cite{gi02}. In fact, we would like to remark that the second
diffraction minimum is in a range of $Q^2$ where the applicability of
the standard non-relativistic nuclear physics approach presented here may be
questionable.

Also shown in Fig. 2 is the asymptotic prediction of the
dimensional-scaling quark model by Brodsky and Chertok~\cite{br76},
arbitrarily normalized at $Q^2$~=~40~fm$^{-2}$.
It is evident that the data rule out conclusively the
applicability of the long-standing quark dimensional scaling prediction
for elastic electron-$^4$He scattering, at least in the $Q^2$ range
accessible by the JLab accelerator.

In summary, we have measured the $^4$He charge form factor
in the range 29 fm$^{-2}$ $\le Q^2 \le 77$ fm$^{-2}$.
The new data have uncovered a second diffraction minimum for this
form factor.  They constrain
inherent uncertainties of the theoretical calculations and lead,
together with previous large $Q^2$ data on the deuteron, $^3$He and tritium
elastic form factors~\cite{al99,ar78,am94}, to the
development of a consistent hadronic model describing the
internal EM structure and dynamics of few-body nuclear systems.

We acknowledge the outstanding support of the staff
of the Accelerator and Physics Divisions of JLab
that made this experiment possible.  We are grateful to
Drs. D.~Riska, R.~Schiavilla and R.~Wiringa
for kindly providing their theoretical calculations, and to Drs. F.~Gross
and R.~Schiavilla for valuable discussions and support.
This work was supported in part by the U.S. Department of Energy
and National Science Foundation, including grants
DE-AC05-060R23177, DE-AC02-06CH11357, DE-FG02-96ER40950,
NSF-PHY-0701679 and NSF-PHY-0652713,
the Kent State University Research Council and
the Italian Institute for Nuclear Research.

\vspace* {-.2in}
{}

\end{document}